\documentclass[universe,conferencereport,accept,oneauthor,pdftex,10pt,a4paper]{mdpi} 
\pdfoutput=1

\firstpage{1} 
\makeatletter 
 \usepackage{xcolor} 
\makeatother
\pubvolume{4}
\issuenum{12}
\articlenumber{136}
\pubyear{2018}
\copyrightyear{2018}
\history{Received: 2 November 2018; Accepted: 26 November 2018; Published: 29 November 2018}
\updates{yes}

\usepackage{amsfonts}
\usepackage{amssymb}
\usepackage{mathabx}
\usepackage{mathrsfs}
\usepackage[english]{babel}
\usepackage{url}
\usepackage[utf8]{inputenc}

\newcommand{\tg}{\tilde{g}}
\newcommand{\tR}{\tilde{R}}
\newcommand{\tG}{\tilde{G}}
\newcommand{\tK}{\tilde{K}}

\newcommand{\bR}{\bar{R}}
\newcommand{\bG}{\bar{G}}
\newcommand{\bK}{\bar{K}}
\newcommand{\bT}{\bar{T}}

\newcommand{\hvf}{\hat{\varphi}}
\newcommand{\hg}{\hat{g}}

\newcommand{\de}{\partial}
\newcommand{\p}{\prime}
\newcommand{\beq}{\begin{equation}}
\newcommand{\eeq}{\end{equation}}

\newcommand{\smid}{\shortmid}

\newcommand{\mcalA}{\mathcal{A}}
\newcommand{\mcalB}{\mathcal{B}}
\newcommand{\mcalC}{\mathcal{C}}

\newcommand{\Msf}{M_{6}^{4}}
\newcommand{\Mft}{M_{5}^{3}}
\newcommand{\Mfs}{M_{4}^{2}}

\Title{Degravitation and the Cascading DGP Model}

\Author{Fulvio Sbis\`a}

\AuthorNames{Fulvio Sbis\`a}

\address[1]{Departamento de F\'isica, Universidade Federal do Esp\'irito Santo, Avenida Fernando Ferrari, 514, CEP~29075-910 Vit\'oria, ES, Brazil; fulviosbisa@gmail.com}

\abstract{We consider the 6D Cascading DGP model, a braneworld model which is a promising candidate to realize the phenomenon of the degravitation of vacuum energy. Focusing on a recently proposed thin limit description of the model, we study solutions where the induced metric on the codimension-2 brane is of the de Sitter form. While these solutions have already been recovered in the literature imposing by hand the bulk to be flat, we show that it is possible to derive them without making this assumption, by solving a suitably chosen subset of the bulk equations.}


\begin{document}

\section{Introduction}

The Cosmological Constant (CC) problem \cite{Zel'dovich:1968, Weinberg:1989} is one of the most striking puzzle in contemporary physics. In a very basic sense, it stems from the fact that in General Relativity (GR) the gravitational field is sensitive to energy and momentum themselves, and not just to the energy difference with respect to a reference (``zero'') configuration. At semi-classical level, where the spacetime is held classical while the matter fields are quantized, the geometry is expected to be sourced by the quantum expectation value of the energy momentum operator, which is badly (quadratically) divergent. While this vacuum energy could be renormalized, at~the cosmological energy scale 10$^{-3}$ eV, to the value inferred by observations, it is not expected to remain small when we consider higher energy descriptions where new particles enter in the effective Lagrangian. In other words, the smallness of the Cosmological Constant is not technically natural (in a Wilsonian sense) \cite{Burgess:2013ara}.

It may however be that the correct question to ask is not why vacuum energy is so small, but instead why it gravitates so little.~The possible existence of extra dimensions opens a new avenue in this direction, allowing vacuum energy being huge while producing substantial effects only in the extra dimensions \cite{Rubakov:1983bz}. Particularly compelling in this sense are the braneworld models \cite{Akama:1982, Rubakov:1983bb}, where our universe is a surface (brane) embedded in a higher dimensional spacetime. These apparently exotic models are motivated by fundamental physics, since extra dimensions and branes are essential ingredients in string theory.

From this perspective, the most promising realisation is that of codimension-2 branes.\footnote{The codimension of a brane is defined as the difference between the dimension of the ambient spacetime and the dimension of the brane.} In~this case, placing vacuum energy ($\Lambda$) on the brane creates a conical singularity in the extra dimensions, while the 4D universe remains flat independently of the value of $\Lambda\,$. To solve the CC problem, a desired feature is that a generic configuration should dynamically relax towards one where the 4D universe is flat, for example after a phase transition changes the value of vacuum energy on the brane. It was shown that this mechanism, called \emph{self-tuning}, cannot work when the extra dimensions are compact, since a phase transition is followed by a runaway behaviour \cite{Vinet:2004bk, Garriga:2004tq}. The reason for this behaviour is intimately linked to the extra dimensions being compactified.

If the extra dimensions have infinite volume, on the other hand, it is necessary to include an induced gravity term on the brane for   gravity to look four dimensional~\cite{Dvali:2000hr}. The~coexistence of these two properties indeed permits   circumventing the Weinberg's no-go theorem \cite{Dvali:2002pe}, and, remarkably, is central also to the \emph{degravitation} approach to the CC problem~\cite{ArkaniHamed:2002fu}, which advocates theories where the strength of the gravitation interaction effectively depends on the characteristic length of the source. To reproduce the successes of GR, a ``high-pass filter'' behaviour is invoked, in such a way that only sources smooth on cosmological scales, such as vacuum energy, get degravitated.

\section{The 6D Cascading DGP Model}

In this paper, we focus on an interesting candidate to realize the self-tuning/degravitation phenomenon, the Cascading DGP model \cite{deRham:2007xp}, originally proposed to extend the DGP model~\cite{Dvali:2000hr} to higher dimensions. In its minimal (6D) set-up, a codimension-2 brane is embedded inside a codimension-1 brane, which in turn is embedded in an infinite-volume 6D spacetime (also called ``bulk'', $\mcalB$). Both branes are equipped with induced gravity terms. The model is schematically described by the action
\beq \label{CascadingDGP6D}
	S = \Msf \int_{\mcalB} \!\! d^{6} X \, \sqrt{-g} \, R + \Mft \int_{\mcalC_1} \!\! d^5 \xi \, \sqrt{-\tg} \, \tR + \int_{\mcalC_2} \!\! d^4 \chi \, \sqrt{-\bar{g}} \, \Big( \Mfs \bR + \mathscr{L}_{_{\!M}} \Big) \quad 
\eeq
where $\tg$ and $\bar{g}$ denote the determinant of the metrics induced respectively on the codimension-1 brane $\mcalC_1$ and on the codimension-2 brane $\mcalC_2$. Moreover, $\tR$ and $\bar{R}$ indicate the associated Ricci scalars, and $\mathscr{L}_{_{\!M}}$ is the Lagrangian for matter, which is localized on the codimension-2 brane (our universe). Quantities pertaining to $\mcalC_1$ are indicated with a tilde, while quantities pertaining to $\mcalC_2$ are indicated with an overbar. While $M_{4}^{-2}$ is fixed to be $8 \pi G/c^{4}$ to reproduce the GR results on small scales, the model has two free parameters $m_{5} = \Mft/\Mfs$ and $m_{6} = \Msf/\Mft$, which control the relative strength of the induced gravity terms and the Einstein--Hilbert term in the bulk.

In the past years, several interesting properties of the model were uncovered, such as a better behaviour concerning ghost instabilities \cite{deRham:2007xp, deRham:2010rw, Sbisa:2014vva} than that of the pure codimension-2 set-up with induced gravity \cite{Dubovsky:2002jm} (see, however, \cite{Berkhahn:2012wg}). Moreover, it was proposed that the codimension-1 brane acts somehow as a regulator, rendering gravity finite at the codimension-2 brane even when its thickness tends to zero \cite{deRham:2007rw, Sbisa:2014gwh}. However, a mathematically clean proof of these properties, as well as the general investigation of exact solutions, were made impossible by the well-known issue that the thin limit of a codimension-2 brane is not well-defined \cite{Geroch:1987qn}. This implies that the action in  Equation  (\ref{CascadingDGP6D}) provides at best a schematic  description, while to properly define the model one should specify in the action the details of the internal structure of the branes. This is in fact a thorny point, since in general there is no detailed notion of how the confining mechanism works, while we are mainly interested in understanding how gravity behaves on scales much larger than the branes thicknesses.

This obstacle has been recently overcome, with the derivation of a thin limit description of the model \cite{Sbisa:2017gsm}, valid for a large class of internal structures. In the thin description, the bulk metric obeys the (6D) Einstein equations
\beq \label{bulk eq}
	G_{_{AB}} = 0
\eeq
while the Israel junction conditions \cite{Israel:1966}
\beq \label{cod1 jc}
	2 \Msf \, \Big( \tK_{ab} - \tK \, \tg_{ab} \Big) + \Mft \, \tG_{ab} = \,\, 0
\eeq
are imposed at the codimension-1 brane, which acts as a boundary for the bulk. The energy momentum-tensor $\bT_{\mu\nu}$ localized on the codimension-2 brane causes the Einstein tensor $\tG_{ab}$ and the extrinsic curvature $\tK_{ab}$ of the codimension-1 brane to be discontinuous at the codimension-2 brane, while leaving smooth the bulk geometry. This discontinuity is encoded in the codimension-2 junction conditions
\beq \label{cod2 jc}
	- \Msf \, \Delta \,\, \bar{g}_{\mu\nu} + 2 \Mft \, \Big( \bK_{\mu\nu} - \bK \, \bar{g}_{\mu\nu} \Big) + \Mfs \, \bG_{\mu\nu} = \,\, \bT_{\mu\nu} \quad 
\eeq
where $\bar{g}_{\mu\nu}$ is the induced metric on the codimension-2 brane; $\bG_{\mu\nu}$ and $\bK_{\mu\nu}$ are, respectively, the Einstein tensor and the extrinsic curvature of the codimension-2 brane; and $\Delta$ is the local deficit angle of the configuration.

\section{De Sitter Solutions}

While the general study of exact cosmological solutions of the system in Equations  (\ref{bulk eq})--(\ref{cod2 jc}) is under way and will be published elsewhere, we want to show here how the self-accelerating solutions found in Ref. \cite{Minamitsuji:2008fz} can be obtained in this description.

The aforementioned solutions are characterized by the bulk being flat and the induced geometry on the codimension-2 brane being de Sitter,
with the codimension-2 brane being empty and the four-dimensional Hubble (constant) factor obeying the equation
\beq \label{Leve Minas}
	\frac{H}{2 \, m_5} - \sqrt{1 - \frac{16 \, m_{6}^{2}}{9 H^{2}} \, } \, + \frac{2 \, m_{6}}{3 H} \, \arctan \sqrt{\frac{9 H^{2}}{16 \, m_{6}^{2}} - 1} = 0 \quad 
\eeq

In Ref. \cite{Sbisa:2017gsm}, Appendix G, we showed that these solutions can be recovered in the thin limit description, imposing by hand the bulk to be flat. Actually, we found a slightly more general class of solutions, describing both contracting and expanding universes in presence of a deficit or excess angle.   Equation (\ref{Leve Minas}) corresponds to the case of an expanding universe in presence of a deficit angle. We want now to obtain these solutions in a more satisfactory way, by~specifying only the symmetries of the spacetime, and solving the equations of motion (bulk included) in a consistent way.

\subsection{Symmetry Properties and Gauge Choice}

Since the source term is highly symmetric (being actually vanishing), we consider the following hypotheses about the bulk geometry:
\begin{enumerate}[leftmargin=*,labelsep=4.9mm]
	\item The bulk can be foliated into 4D leaves, whose (pseudo-Riemannian) induced metric is spatially homogeneous and isotropic.
	\item The translations along the 3D spatial dimensions are a symmetry of the bulk metric.
	\item There exists a smooth hypersurface $\Sigma$ with respect to which the bulk is $\mathbb{Z}_{2}$-symmetric. The codimension-2 brane coincides with the intersection of $\Sigma$ and the codimension-1 brane. 
	\item The $\mathbb{Z}_{2}$-symmetry and the foliation into spatially homogeneous and isotropic leaves are compatible.
\end{enumerate}

Point 3 ~in particular implies that the total six-dimensional configuration is $\mathbb{Z}_2 \times \mathbb{Z}_2$ symmetric. This~is not a general property in the formulation of the Cascading DGP model given in Ref.~\cite{Sbisa:2017gsm}, but it is an assumption which seems sensible for highly symmetric configurations such as the one we are considering. 

\subsubsection{The Bulk}

It is convenient to use the gauge freedom to reduce the number of unknown functions. Considering Gaussian Normal Coordinates with respect to $\Sigma\,$, and calling $y$ the coordinate normal to the hypersurface, we obtain the following bulk line element
\beq \label{bulk line}
	ds^2 = -N^{2}(t,y,z) \, dt^2 + A^{2}(t,y,z) \, \delta_{ij} \, dx^i dx^j + 2 \, F(t,y,z) \, dt \, dz + dy^2 + B^{2}(t,y,z) \, dz^2 \quad 
\eeq
where $\Sigma$ is identified by $y = 0$ and $N$, $A$, $F$ and $B$ are even with respect to $y$. In particular, the~last point implies that
\beq \label{Arcimboldi}
	\de_{y} N \big\rvert_{y = 0} = \de_{y} A \big\rvert_{y = 0} = \de_{y} F \big\rvert_{y = 0} = \de_{y} B \big\rvert_{y = 0} = 0 \quad 
\eeq

We can furthermore take advantage of the bulk gauge freedom associated to the coordinate choice on $\Sigma\,$. Considering, inside $\Sigma\,$, Gaussian Normal coordinates with respect to the codimension-2 brane, the latter is identified by $z = 0$ and we have
\begin{align} \label{y=0 gauge fix}
	B(t,0,z) = 1, \quad ~~~~~~~~~~~~~~F(t,0,z) = 0 \quad 
\end{align}

\begin{figure}[H]
\centering
\begin{tikzpicture}
\begin{scope}[scale=.4,>=stealth]
\fill[color=green!20!white] (0,0) -- ++(80:6) -- ++(10:14) node[color=black,anchor=north east]{$\mcalC_1$} -- ++(260:6);
\fill[color=green!20!white] (0,0) -- ++(280:6) -- ++(10:14) node[color=black,anchor=south east]{$\mcalC_1$} -- ++(100:6);
\draw (0,0) -- ++(80:6) -- ++(10:14) -- ++(260:6);
\draw (0,0) -- ++(280:6) -- ++(10:14) -- ++(100:6);
\draw[very thick] (0,0) -- +(10:14);
\draw[dashed] (0,0) -- (18,0) -- ++(10:14) node[color=black,anchor=south east]{$\Sigma$} -- ++(180:18);
\draw[thin,color=red,->] (5,4) node[color=black,anchor=east]{$\mcalC_2$} -- (9,1.65);
\draw[thin,->] (16.5,1) -- (16.5,4) node[color=black,anchor=south]{$y$};
\draw[thin,->] (16.5,1) -- (19.5,1) node[color=black,anchor=west]{$z$};
\draw[thin,->] (16.5,1) -- ++(10:2) node[color=black,anchor=south]{$t$};
\end{scope}
\end{tikzpicture}
\caption{Pictorial representation of the configuration and of the coordinate choice.}
\label{figure 1}
\end{figure}
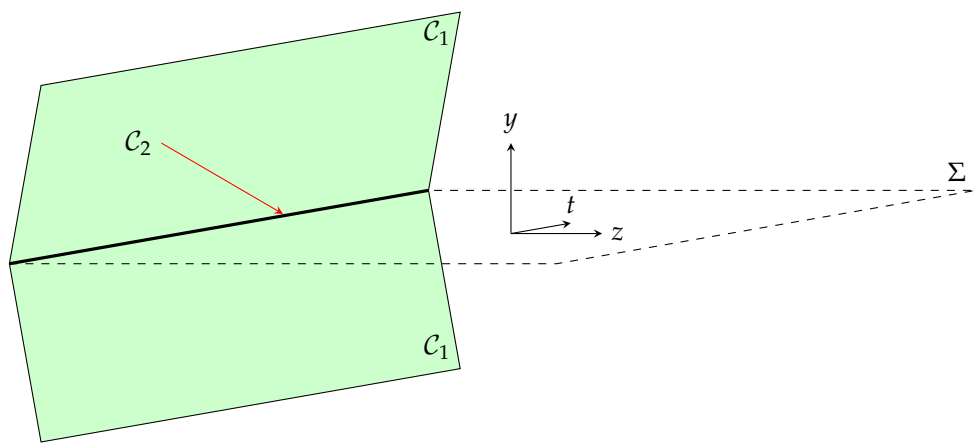

Therefore, naively speaking, $y$ is the coordinate orthogonal both to the codimension-2 brane and to $\Sigma \,$, while $z$ is the coordinate orthogonal to the codimension-2 brane and parallel to $\Sigma\,$. The coordinates $t$, $x^{1}$, $x^{2}$ and $x^{3}$ are instead parallel to the codimension-2 brane. A sketch of the configuration and of the coordinate choice is given in Figure \ref{figure 1}.

We can finally use the bulk gauge freedom associated to the coordinate choice on the codimension-2 brane, which is now placed at $y = z = 0\,$. Rescaling the time coordinate in a suitable way we can set
\beq \label{gauge fix N}
	N(t,0,0) = 1 \quad 
\eeq
so that the induced line element on the codimension-2 brane reads
\beq \label{cod2 induced metric}
	d \bar{s}^{2} = - dt^{2} + a^{2}(t) \, \delta_{ij} \, dx^{i} dx^{j} \quad 
\eeq
where we defined
\beq \label{scale factor}
	a(t) \equiv A(t,0,0) \quad 
\eeq

\subsubsection{The Branes}

Regarding the gauge choice for the branes, we closely follow   Appendix G of Ref.~\cite{Sbisa:2017gsm}. On the codimension-1 brane we choose Gaussian Normal coordinates with respect to the codimension-2 brane, thus, indicating with $\zeta$ the coordinate orthogonal to the codimension-2 brane, we have
\begin{align} \label{cod1 GNC}
	\hg_{\zeta\zeta}(t\, , \vec{x}\, , \zeta) = 1,~~~~~~~~~~~~~~~~\hg_{\zeta\mu}(t\, , \vec{x}\, , \zeta) = 0 \quad 
\end{align}
where we adopt the notation of indicating codimension-1 quantities evaluated in this coordinate system with an overhat. In this coordinate system, the codimension-2 brane therefore lies at $\zeta = 0\,$, and we assume that the codimension-1 embedding is of the form
\beq \label{cod1 embedding}
	\hvf^{_{A}}(t\, , \vec{x}\, , \zeta) = \Big( t\, , \vec{x}\, , \hvf^{y}(t, \zeta) , \hvf^{z}(t, \zeta) \Big) \quad 
\eeq

Since, in the bulk coordinates, the codimension-2 brane is placed at $y = z = 0\,$, by consistency it follows that
\beq \label{oppaaa}
	\hvf^{y}(t\, , 0) = 0 \qquad \quad \qquad \hvf^{z}(t\, , 0) = 0 \quad 
\eeq

As we showed in Ref. \cite{Sbisa:2017gsm}, the condition in Equation (\ref{cod1 GNC}) implies that the derivatives $\de_{\zeta} \hvf^{y}$ and $\de_{\zeta} \hvf^{z}$ at the $\zeta = 0^{+}$ side of the codimension-2 brane can be written in terms of the local deficit angle as follows
\begin{align}
	\de_{\zeta}\Big\rvert_{0^{+}} \hvf^{y}(t) &= \cos \frac{\Delta(t)}{4} & \de_{\zeta}\Big\rvert_{0^{+}} \hvf^{z}(t)&= \sin \frac{\Delta(t)}{4} \quad 
\end{align}

\subsection{Degrees of Freedom and Flat Bulk Configurations}

It is useful at this point to recognise which are the degrees of freedom that we are free to assign to individuate the six-dimensional configuration in a neighbourhood of the codimension-2 brane. Since the equations of motion are differential equations of second order, the degrees of freedom are to be found among the embedding functions of the branes and the bulk metric components, and their first derivatives.

Since the position of the codimension-2 brane is held fixed, both inside the codimension-1 brane and in the bulk, the only degrees of freedom associated to the codimension-2 brane are the scale factor $a(t)$ and the Hubble factor $H(t) = \dot{a}/a\,$. The configuration of the codimension-1 embedding in a neighbourhood of $\zeta = 0\,$, on the other hand, is determined by the local deficit angle $\Delta (t)$ only. Finally, the bulk configuration in a neighbourhood of $y = z = 0$ is determined by the derivatives $\de_{z} A$ and $\de_{z} N\,$, since all   other first order partial derivatives at $y = z = 0$ vanish (by symmetry or by gauge choice). For future convenience, it is preferable to work with the quantity $\pi(t) = \de_{z} A(t,0,0)/a$ instead of with $\de_{z} A (t,0,0)\,$, and we indicate $N_{\smid}(t) = \de_{z} N (t,0,0) \,$. Since the scale factor $a(t)$ can be rescaled without affecting the equations of motion, we conclude that the degrees of freedom of the six-dimensional configuration in a neighbourhood of the codimension-2 brane are $H(t)\,$, $\Delta (t)\,$, $\pi(t)\,$ and $N_{\smid}(t)\,$.

Let us now identify which conditions on the degrees of freedom are imposed by the request of the bulk geometry being flat. Note that the flat bulk configurations considered in Ref.~\cite{Sbisa:2017gsm}, Appendix G are a special subclass of the configurations in Equation (\ref{bulk line}), and are in fact those for~which
\begin{align} \label{flat bulk}
	N(t,y,z) &= 1 + \epsilon \,  z \, \frac{\ddot{a}}{\dot{a}} \quad , & A(t,y,z) &= a + \epsilon \, z \, \dot{a} \quad , & F(t,y,z) &= 0 \quad , & B(t,y,z) &= 1 \quad 
\end{align}
where $\epsilon = \pm 1$ and an overdot indicates derivation with respect to time. Incidentally, these~expressions are compatible with our gauge choices. In relation to the discussion of the previous paragraph, we conclude that the flat bulk configurations are those for which the degrees of freedom $H$, $\pi$ and $N_{\smid}$ are linked by the relations
\begin{align} \label{flat bulk and dof}
	\pi(t) = \epsilon \, H ,\quad~~~~~~~~~~N_{\smid}(t) = \pi + \frac{\dot{\pi}}{H} \quad 
\end{align}

If we restrict our attention to the de Sitter solutions, then these relations reduce to
\begin{align} \label{flat bulk and dof de Sitter}
	\pi = \epsilon \, H ,\quad~~~~~~~~~~~N_{\smid} &= \pi \quad 
\end{align}

\section{The Role of the Bulk Equations}

It is worthwhile at this point to recall some relevant points about the analysis of Ref.~\cite{Sbisa:2017gsm}, Appendix G. The de Sitter solutions were found by using the codimension-2 junction conditions and the $\zeta\zeta$ and $\zeta\mu$ components of the codimension-1 (Israel) junction conditions evaluated at $\zeta = 0\,$. The $\mu\nu$ components of the latter were not used since they are evolution equations for the codimension-1 embedding in the direction normal to the codimension-2 brane, and~we were interested only on the behaviour of the Hubble factor (i.e.,\ on what happens on the codimension-2 brane). For the same reason, the codimension-1 junction conditions away from $\zeta = 0$ were not used. Remarkably, imposing the conditions in Equation  (\ref{flat bulk}) which ensure the bulk flatness, this restricted set of equations permits to determine completely the solutions for the induced metric on the codimension-2 brane, and in particular to derive   Equation (\ref{Leve Minas}) for the solutions with $H$ constant.

On the other hand, if the thin limit formulation in Equations (\ref{bulk eq})--(\ref{cod2 jc}) of the model is consistent, one~should be able to determine the solution on the codimension-2 brane without imposing any condition on the bulk. That is, it should be possible to find the solution by specifying the source content on the codimension-2 brane, solving (all) the equations of motion and imposing boundary conditions at infinity in the extra dimensions. This is in general a formidable task, due to the complexity of the equations. However, we show now that, in the highly symmetric case we are considering, it is indeed possible to derive the relations in Equation  (\ref{flat bulk and dof}) by solving the bulk equations, without making strong a priori hypothesis on the bulk geometry.

\subsection{Initial Value Analysis}
\label{Initial value analysis}

The main idea of the analysis is to study the bulk equations from the point of view of the initial value formulation with respect to the hypersurface $\Sigma \,$. It is well-known that solutions of the bulk Einstein equations exist in a neighbourhood of $\Sigma$ if the following constraint equations are satisfied
\begin{align} \label{Sigma bulk constraints}
	G_{yy}\big\rvert_{\Sigma} = 0~~~~~~~~~~~G_{yz}\big\rvert_{\Sigma} = 0~~~~~~~~~~~~G_{y\mu}\big\rvert_{\Sigma} = 0 \quad 
\end{align}

Any choice of the induced metric on $\Sigma$ and of $\Sigma$'s extrinsic curvature determine a unique local solution of the bulk Einstein equations, provided the relations in  Equation (\ref{Sigma bulk constraints}) are satisfied. It is not difficult to see that, with our choice of the bulk metric, the equations $G_{yz}\rvert_{_{\Sigma}} = 0$ and $G_{y\mu}\rvert_{_{\Sigma}} = 0$ are identically satisfied. Therefore, the only non-trivial bulk constraint equation is $G_{yy}\rvert_{_{\Sigma}} = 0\,$, to wit
\beq \label{bulk constraint eq}
	\frac{(\de_{z} A)^{2}}{A^{2}} + \frac{\de_{z} A \, \de_{z} N}{A N} + \frac{\de_{z}^{2} A}{A} + \frac{\de_{z}^{2} N}{3 N} - \frac{\dot{A}^{2}}{A^{2} N^{2}} + \frac{\dot{A} \dot{N}}{A N^3} - \frac{\ddot{A}}{A N^{2}} = 0 \quad 
\eeq
where it is understood that $A = A(t, 0, z)$ and $N = N(t, 0, z)\,$. We refer to this equation as the \emph{bulk constraint equation}. Regarding the other components of the bulk equations, it is important to notice that the equation $G^{\, t}_{\,\,\, z} = 0$ is the evolution equation for the off-diagonal metric component $F$, since
\beq \label{bulk mixed Einstein B=1}
	G^{\, t}_{\,\,\, z}\big\rvert_{\Sigma} = \frac{3}{N^{2}} \, \bigg( \frac{\de_{y}^{2} \, F}{6} + \frac{\de_{z} \dot{A}}{A} - \frac{\dot{A} \, \de_{z} N}{A N} \bigg) \quad 
\eeq

Reminding that $F\vert_{_{\Sigma}} = 0$ by gauge choice and $\de_{y} F\vert_{_{\Sigma}} = 0$ by symmetry, it follows that the bulk metric is diagonal if and only if the following \emph{diagonality condition} is satisfied
\beq \label{diagonality condition}
	\bigg[ \, \frac{\de_{z} \dot{A}}{A} - \frac{\dot{A} \, \de_{z} N}{A N} \, \bigg]_{\Sigma} = 0 \quad ,
\eeq
since by  Equation (\ref{bulk mixed Einstein B=1}) it implies $\de_{y}^{2} \, F\vert_{_{\Sigma}} = 0\,$.

\subsection{Diagonal Solutions}
\label{Diagonal solutions}

Let us specialize to the configurations where the bulk metric in Equation  (\ref{bulk line}) is diagonal. In this case, the diagonality condition in  Equation \eqref{diagonality condition} can be integrated to give
\beq
\label{Mariana}
N(t, z) = \frac{\dot{A}(t, z)}{\dot{a}(t)} \quad 
\eeq
and defining for convenience
\beq
\mcalA(t,z) = \frac{A(t,z)}{a(t)} \quad 
\eeq
the bulk constraint in Equation (\ref{bulk constraint eq}) can be written as
\beq \label{Daiane}
	\frac{\de_{z}^{2} \mcalA}{\mcalA} + \frac{1}{3} \, \frac{\de_{z}^{2} \dot{\mcalA} + H \, \de_{z}^{2} \mcalA}{\dot{\mcalA} + H \mcalA} + \frac{\de_{z} \mcalA}{\mcalA} \bigg( \frac{\de_{z} \mcalA}{\mcalA} + \frac{\de_{z} \dot{\mcalA} + H \, \de_{z} \mcalA}{\dot{\mcalA} + H \mcalA} \bigg) - \frac{H}{\mcalA} \bigg( \frac{H}{\mcalA} + \frac{\dot{H} + H^{2}}{\dot{\mcalA} + H \mcalA} \bigg) = 0 \quad 
\eeq

Note that, by definition, the quantity $\mcalA$ on the codimension-2 brane obeys
\begin{align} \label{boundary conditions cod2 brane}
	\mcalA(t,0) = 1~~~~~~~~~~~~~~~\de_{z} \mcalA(t,0) = \pi(t) \quad 
\end{align}

From what is said above, it follows that a unique solution of the bulk equations in a neighbourhood of $\Sigma$ is associated to each solution of   Equation (\ref{Daiane}) which satisfies the boundary conditions in Equation  (\ref{boundary conditions cod2 brane}) at the codimension-2 brane and suitable boundary conditions at infinity in the extra dimensions.

Regarding the latter boundary conditions, it is usually assumed that the gravitational field decays at infinity in the extra dimensions, and that eventual radiation is purely outgoing (``nothing enters from infinity in the extra dimensions''). Since we are not considering gravitational wave solutions, we therefore impose the bulk Riemann tensor to tend to zero when $y$ and/or $z$ tend to infinity. Remarkably, this requirement induces a restricted set of boundary conditions at infinity on $\Sigma$ (i.e.,~at $z \to \infty$ with $y = 0$), as a consequence of the fact that the components
\beq \label{Riemann components}
	R_{titi}\big\rvert_{_{\Sigma}} \quad , \quad R_{tiiz}\big\rvert_{_{\Sigma}} \quad , \quad R_{tztz}\big\rvert_{_{\Sigma}} \quad , \quad R_{ijij}\big\rvert_{_{\Sigma}} \quad , \quad R_{iziz}\big\rvert_{_{\Sigma}}
\eeq
involve only $A\,$, $N$ and their derivatives with respect to $t$ and $z$ (since Equation  (\ref{Arcimboldi}) holds). Asking the components in Equation  (\ref{Riemann components}) of the Riemann tensor at $\Sigma$ to decay to zero when $z \to \infty$ gives the following conditions
\begin{align} \label{boundary conditions infinity diagonal}
	(\de_{z} \mcalA)^{2} \xrightarrow[z \to \infty]{} H^{2}~~~~~~~~~~~~~~\de_{z}^{2} \mcalA \xrightarrow[z \to \infty]{} 0 \quad 
\end{align}

\subsection{De Sitter Solutions Again}

Summing up, in Sections \ref{Initial value analysis} and \ref{Diagonal solutions}, we show  that, restricting the analysis to diagonal configurations, a unique solution of the Einstein equations in the bulk is associated to each solution of   Equation (\ref{Daiane}) subject to the boundary conditions in Equation  (\ref{boundary conditions cod2 brane}) at the codimension-2 brane and to the boundary conditions in Equation (\ref{boundary conditions infinity diagonal}) at infinity on $\Sigma\,$.

The solution of such a problem is indeed very challenging, since for one thing   Equation~(\ref{Daiane}) is non-linear. However, as long as we are interested in the de Sitter solutions, we can make the simplifying assumption $\dot{H} = \dot{\pi} = 0\,$. Taking into account the boundary conditions in Equations  (\ref{boundary conditions cod2 brane}) and (\ref{boundary conditions infinity diagonal}), this implies that $\mcalA$ and its $z$-derivative are independent of time at both boundaries $z = 0$ and $z \to \infty\,$. Since the bulk equation is of second order in the derivatives with respect to $z\,$, it is natural to assume that $\mcalA$ only depends on $z\,$. In this case, indicating derivatives with respect to $z$ with a prime, the bulk in Equation (\ref{Daiane}) becomes the ordinary differential equation
\beq \label{bulk De Sitter eq}
	\frac{2}{3} \, \frac{\mcalA^{\p\p}}{\mcalA} + \frac{\mcalA^{\p \, 2}}{\mcalA^{2}} - \frac{H^{2}}{\mcalA^{2}} = 0 \quad 
\eeq
and a solution compatible with the boundary conditions in Equation (\ref{boundary conditions infinity diagonal}) at infinity is clearly given by
\begin{align}
	\mcalA^{\p\p} (z) = 0~~~~~~~~~~~~~~~~~~~\mcalA^{\p \, 2} (z) = H^{2} \quad 
\end{align}

Absorbing the sign ambiguity by introducing a constant $\epsilon = \pm 1\,$, we get $\mcalA^{\p} = \epsilon \, H\,$. Taking~then into account the boundary conditions at the codimension-2 brane in Equation (\ref{boundary conditions cod2 brane}), we get
\begin{align} \label{solution mcalA pi}
	\mcalA(z) = 1 + \epsilon \, z \, H, \quad~~~~~~~~~~~~~~\pi &= \epsilon \, H \quad 
\end{align}
and evaluating the diagonality condition in Equation (\ref{diagonality condition}) at $y = z = 0$, we find
\beq
N_{\smid} = \pi \quad
\eeq

We then reproduced the conditions in Equation (\ref{flat bulk and dof de Sitter}), which were originally obtained by imposing by hand the bulk to be flat, by solving the bulk equations and imposing suitable boundary conditions at infinity.

An important question is whether  Equation (\ref{solution mcalA pi}) is the \emph{only} solution of   Equation (\ref{bulk De Sitter eq}), and more in general of  Equation (\ref{Daiane}), compatible with the boundary conditions in Equations (\ref{boundary conditions cod2 brane}) and (\ref{boundary conditions infinity diagonal}) when $\dot{H}~=~\dot{\pi}~=~0\,$. In~particular, the argument leading to the assumption of $\mcalA$ being time-independent is somehow qualitative, and a thorough analysis of the boundary condition problem given by   Equation~(\ref{Daiane}) and the conditions in Equations (\ref{boundary conditions cod2 brane}) and (\ref{boundary conditions infinity diagonal}) would be needed to justify it properly. We have no definitive answer to these questions at present, but we aim to consider the matter in detail in a forthcoming publication.

\section{Conclusions}

We considered the 6D Cascading DGP model, a braneworld model which is a promising candidate to realize the phenomenon of the degravitation of vacuum energy. Our main objective was to show that the self-accelerating (de Sitter) solutions of Ref. \cite{Minamitsuji:2008fz} can be obtained in the thin limit formulation of the model recently derived in Ref. \cite{Sbisa:2017gsm}, without imposing by hand the bulk to be flat. Making suitable assumptions on the symmetry properties of the bulk, and~restricting the analysis to configurations where the bulk metric is diagonal, we identified a component of the bulk Einstein equations which can be naturally and consistently equipped with boundary conditions at the codimension-2 brane and at infinity in the extra dimensions. We showed that, when the Hubble factor on the codimension-2 brane is time independent, solving the aforementioned bulk equation indeed permits   obtaining the de Sitter solutions without a priori imposing the bulk to be flat. A natural extension of this analysis would be the study of diagonal cosmological solutions in this set-up. Work in this direction is under way.


\vspace{6pt} 

\funding{This research received no external funding.}

\acknowledgments{The author acknowledges partial financial support from CNPq (Brazil) and FAPES (Brazil).}

\conflictsofinterest{The author declares no conflict of interest.} 


\reftitle{References}

\end{document}